\documentclass[12pt,letterpaper]{JHEP3}
\usepackage{graphicx}
\newcommand{\eq}[1]{Eq.~(\ref{#1})}
\newcommand{\eqs}[2]{Eqs.~(\ref{#1}) and (\ref{#2})}
\def\be{\begin{equation}}
\def\ee{\end{equation}}
\def\ba{\begin{eqnarray}}
\def\ea{\end{eqnarray}}
\def\half{\frac{1}{2}}
\def\talpha{{\tilde{\alpha}}}
\def\tp{{\tilde{p}}}
\def\Lfour{{}^{(4)}{\cal L}}
\def\tM{\widetilde{M}}
\def\tE{\widetilde{E}}
\title{Electroweak symmetry breaking as a proximity effect}
\author{Sergei Khlebnikov \\
Department of Physics, Purdue University, West Lafayette, IN 47907, USA
}
\abstract{
The proximity effect in condensed matter physics is a mechanism that naturally
produces weak superconductivity. We argue that a braneworld can similarly produce 
a low-energy breaking of the electroweak symmetry, provided that in addition
to the ``normal'' region, occupied by the conventional phase of QCD, there is
a bulk region where the color is in an anisotropic (layered) state 
with a larger confinement scale. The
$W$ and $Z$ bosons, as well as the quarks, acquire masses by scattering off
the layered region. A peculiar feature of this
scenario is that the strongly interacting sector responsible for the 
symmetry breaking can be much lighter than the conventional 1 TeV.
}
\preprint{}
\begin{document}
\section{Introduction}
The proximity effect in condensed matter physics is a mechanism
by which a material becomes  weakly superconducting when placed
in the proximity of a conventional (``strong'') 
superconductor. Given the parallel between
superconductivity and the electroweak symmetry breaking, 
one may wonder if a similar mechanism is at work also in the latter case.

A natural framework for exploring this possibility is anisotropic states
of higher-dimensional lattice gauge theories. Perhaps the best known of these
is the layer phase 
\cite{Fu:1983ei,Fu:1984gj,Berman:1985qj,Hulsebos:1994pa,Dimopoulos:2000ej,Dimopoulos:2006qz}, 
in which the gauge field is localized in $d$-dimensional layers, so that
the charges within each layer interact
according to the Coulomb law. The requirement that the Coulomb phase be present 
is the reason why for $d=4$ the genuine 
layer phase can exist only in Abelian gauge theories \cite{Fu:1983ei}.

For our purposes, however, the Coulomb law is unnecessary and,
indeed, detrimental. We are interested in the situation when 
the layers are confining, but the confinement (correlation) length
within the layers is much larger than in the direction perpendicular to them.
We do not require such states to form a separate phase; in fact, we expect 
them to be a corner of the usual confining phase of the higher-dimensional 
theory.

We refer to such a state as a layered state or 
a layered region, as opposed to a layer phase.
In Sec.~\ref{sec:struc}, we argue that for the ${\rm SU}(3)$ gauge group in $d+1=5$ 
dimensions it is separated from the weakly-coupled 
phase by a first-order phase transition.

Additional possibilities appear when the fifth dimension is appreciably curved,
as in the case, for example, 
of a brane bounding space with a negative cosmological constant
\cite{Randall:1999vf}.\footnote{
The layer phase of the ${\rm U}(1)$ gauge theory on such a background was
studied in Ref.~\cite{Dimopoulos:2000ej}.} For ${\rm SU}(3)$, the resulting variation 
of the gauge coupling produces a weakly-coupled ``normal''
region near the brane and a layered region further away, with a well defined phase
boundary between them. Because the length of
the ``normal'' region (indeed, in our case, of the entire fifth dimension) 
is finite, we expect that at large 4-dimensional distances the 
5-dimensional theory there reduces to its 4-dimensional counterpart.
So, in this limit, both the ``normal'' and layered regions are
confining, but with different confinement scales: in the ``normal''region, the
confinement scale is the usual $\Lambda_{\rm QCD}$, while in the layers it is
a much larger $\Lambda$.  We suggest that the 
quark condensates of the layered ${\rm SU}(3)$ color are the origin of the masses 
of the $W$ and $Z$ bosons.

The argument of Sec.~\ref{sec:struc} is based on the mean-field theory 
\cite{Balian:1974ts}, which had been a reliable guide to 
the phase structure of anisotropic 5-dimensional Abelian theories
\cite{Fu:1983ei,Fu:1984gj,Berman:1985qj,Hulsebos:1994pa,Dimopoulos:2000ej,Dimopoulos:2006qz}.
At the mean-field level, the ``normal'' phase is ordered, with respect to the link
variable connecting individual 4-dimensional hyperplanes. In
the non-Abelian case (with a finite fifth dimension), we expect that this order is
destroyed by confinement, and in this sense the mean field is formally in error.
However, if the mean-field phase
transition is strongly first-order, and the confinement length is large, we 
do not expect
the transition to disappear altogether. This is the rationale on which we proceed 
here.

Instead of a full 5-dimensional lattice, one could use deconstruction 
\cite{ArkaniHamed:2001ca,Hill:2000mu}---the approach where one 
latticizes only the additional, fifth dimension or, in other words, considers 
multiple instances of the 4-dimensional theory connected to one another by 
sigma-model variables. These sigma-model variables correspond to the link 
variables $U_\talpha$ of the lattice gauge theory ($\talpha$ labels the links).
The simplest version of the proximity effect 
is realized when there is a first-order phase transition in the sigma model.
If, as in the present case, 
$U_\talpha$ are elementary ${\rm SU}(3)$ fields living on the
sites of a 4-dimensional lattice, there is evidence for such a transition already 
at the mean-field level \cite{Kogut:1981ez}. We expect very similar physics in the
case when the nonlinear sigma model is replaced with a linear one
(i.e., $U_\talpha^\dagger U_\talpha$ is allowed to fluctuate), as in one version 
of deconstruction 
\cite{ArkaniHamed:2001ca,Hill:2000mu}. In another version \cite{ArkaniHamed:2001ca},
$U_\talpha$ are composites of new confining 
gauge theories. To explain quark masses, we require that there are quark hopping 
terms, of the form $\bar{q}_\alpha U_\talpha q_{\alpha + 1}$. The presence of such 
terms suggests that,
if $U_\talpha$ are ``mesons'' of a new confining theory, the quarks must be 
``baryons,'' meaning that they are also composite. This is a fascinating 
possibility, but also one more challenging to explore than the fully latticized
theory considered here.

Our proposal can also be compared to Higgsless models 
\cite{Csaki:2003dt,Csaki:2003zu}, 
where the electroweak symmetry is broken by boundary conditions 
in the extra dimension. Unlike the boundary conditions, the layered-state 
quark condensates deform the profiles of $W$ and $Z$ relatively little,
provided the fifth dimension is sufficiently short. (These profiles are 
discussed further in Sec.~\ref{sec:masses}.) As a result, no new
gauge symmetries are required to avoid a large tree-level contribution to 
the ratio of the $W$ and $Z$ masses (the $\rho$ parameter).

Overall, we find that the layered-state color 
functions much like the conventional technicolor 
\cite{Weinberg:1979bn,Susskind:1978ms,Farhi:1980xs}, with the following differences.

(i) The confinement scale $\Lambda$ of the layer theories does not have to be of order
$v \sim 250$ GeV: because more than one layer can contribute to the $W$ mass, it
can be much smaller than that. The masses of various new particles produced by the
strongly coupled sector are then much 
smaller than the conventional 1 TeV. TeV-scale masses of new particles are a major
reason why the conventional technicolor runs into conflict with the measured value 
of the $S$ parameter \cite{Peskin:1990zt}. A lower confinement 
scale may improve prospects for obtaining an acceptable value 
(although we have not yet ascertained whether it actually does so).

(ii) Quarks acquire masses by scattering off the layered region (a process 
similar to Andreev reflection \cite{Andreev}
in a superconductor-normal-superconductor junction). So,
the problem of quark masses, which is quite formidable in the conventional
technicolor, has a natural solution here.

(iii) The quark scattering off the layers can break flavor symmetries. 
This may help to lift 
the masses of pseudo-Goldstone bosons produced by chiral symmetry breaking in 
the layers to an acceptable level.

\section{Layered states in five-dimensional gauge theories}
\label{sec:struc}

Consider a background of the Randall-Sundrum type \cite{Randall:1999vf}, with
the line element
\be
ds^2 = dz^2 + a^2(z) \eta_{\mu\nu} dx^\mu dx^\nu \equiv g_{MN} dx^M dx^N \, .
\label{met}
\ee
Greek indices run from 0 to 3, and $\eta_{\mu\nu}=\mbox{diag}(-1,1,1,1)$ 
is the Minkowski metric tensor. Similarly to Ref.~\cite{Randall:1999vf},
we consider an orbifolded $z$ direction, $z \in S^1/Z_2$, with two branes
at the orbifold's fixed points: one, near which we live, at $z=0$, where 
the warp factor $a(z)$ is maximal, and the other at
$z=\pi R$. We do not, however, take $R$ to infinity, so the fifth dimension 
remains compact. A gauge field $A_\mu$ is assumed even under the orbifold
group $Z_2$, $A_5$ odd; the orbifold projection of fermions will be defined 
in Sec.~\ref{sec:ferm}.

The naive continuum action of a 5-dimensional gauge theory is
\be
S_{\rm cont} =
- \frac{1}{4g_5^2} \int d^4 x dz \sqrt{-g} g^{\mu\nu} \left( g^{\rho\sigma}
G^a_{\mu\rho} G^a_{\nu\sigma}
+ 2 g^{zz} G^a_{\mu z} G^a_{\nu z} \right) ,
\label{cont}
\ee
where $G^a_{\mu\rho}$ and $G^a_{\mu z}$ are the field strengths.
The two terms in \eq{cont} contain different
metric functions, which leads to an anisotropy of the gauge coupling.

There is a degree of arbitrariness in choosing the lattice 
discretization of \eq{cont}. The choice matters: at large $z$, where the theory
is in the layered state, the
4-dimensional hyperplanes decouple from each other, so we are dealing with 
a collection of individual 4-dimensional gauge theories. 
The inverse coupling constants of these theories are
\be
\frac{1}{g_4^2} = \frac{\Delta z}{g_5^2}  \; ,
\label{g4S}
\ee
where $\Delta z$ is the lattice spacing in the $z$ direction.\footnote{
We may need to adjust $\Delta z$ in \eq{g4S} into some effective quantity to take
into account the residual correlations between neighboring hyperplanes. We
retain the notation $\Delta z$ for such a quantity.}
Here we discretize the imaginary-time version of \eq{cont} in such a way that both
$\Delta z$ and $\Delta x$, the spacing in the $x^\mu$ directions,
are independent
of $z$. In this case, $g_4$ and the confinement scale $\Lambda$ of the
4-dimensional theories are also $z$-independent. We assume that
$\Lambda \ll 1/ \Delta x$, so that the layer theories are in the continuum
limit. Then, at one loop,
\be
\Lambda = \frac{1}{\Delta x} 
\exp\left\{ - \frac{b}{g_5^2} \Delta z \right\} \; ,
\label{lam}
\ee
where $b$ is the coefficient of the one-loop beta-function. 

In contrast to $g_4$, the usual coupling constant of QCD, $g_{\rm QCD}$, 
is determined
by the ``normal'' region near the brane (see Fig.~\ref{fig:hyper}):
\be
\frac{1}{g_{\rm QCD}^2} =  \frac{z_{\rm eff}}{g_5^2}  \; ,
\label{gQCD}
\ee
where $z_{\rm eff}$ is the effective 
length of that region (it may be different from
the geometrical length due to effects at the phase boundary).
If $z_{\rm eff}$ is sufficiently large in comparison with $\Delta z$,
the confinement scale (\ref{lam}) is much
larger than $\Lambda_{\rm QCD}$. So, when quarks are added, the chiral
symmetry breaking in the layers will be correspondingly stronger.
This is the basis of the proximity-effect scenario.

\FIGURE[ht]{ \label{fig:hyper}
\includegraphics[scale=0.55]{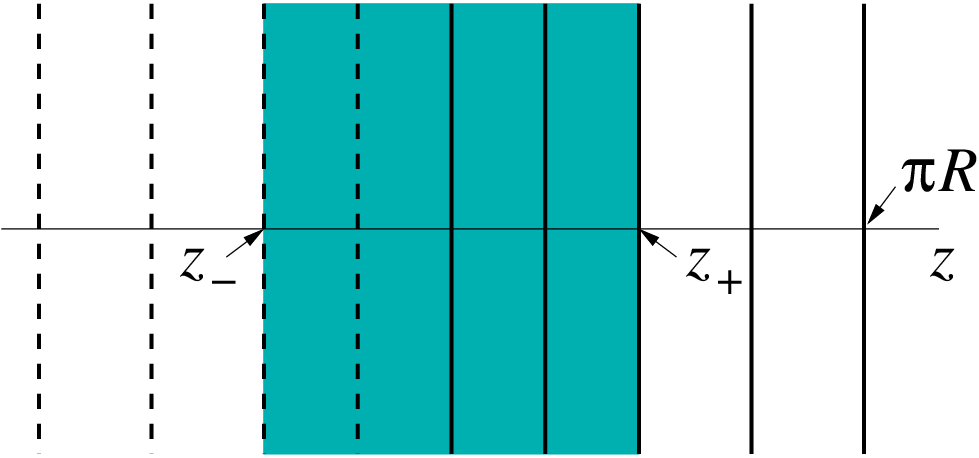}
\caption{Arrangement of 4-dimensional hyperplanes along the $z$ direction. 
The dashed lines are hyperplanes obtained upon ``unfolding'' the orbifold, 
a convenience in the discussion of fermions in Sec.~\ref{sec:ferm}.
The phase boundaries are at $z_{\pm}$. The shaded region is ``normal'';
the rest of the hyperplanes are in the layered state.}
}

The imaginary-time lattice action corresponding 
to \eq{cont} is
\be
S = - \sum_{\alpha,p} \beta_\alpha \chi(U_{\alpha,p}) 
- \sum_{\talpha,\tp} \beta'_\talpha \chi(U_{\talpha,\tp}) \; .
\label{S}
\ee
The integer $\alpha$ labels different (4-dimensional) hyperplanes perpendicular 
to the $z$ axis, $\talpha$ labels the interval between $\alpha$ and $\alpha+1$,
$p$ are plaquettes within a hyperplane, $\tp$ are plaquettes 
connecting the hyperplanes; $\chi(U) = {\rm Tr}U$ for ${\rm SU}(n)$ (where
the trace is taken in the fundamental representation), 
and $\chi(U) = \mbox{Re} U$ for ${\rm U}(1)$.

To simplify notation,
we first write down the mean-field equations for the Abelian group ${\rm U}(1)$.
These are similar to the equations obtained in Ref.~\cite{Balian:1974ts} for
the isotropic Abelian theory.
Once the physics responsible for formation of the layer phase becomes
apparent, we switch back to the non-Abelian case.

We seek to minimize the mean-field energy, whose 4-dimensional density 
(the potential) in the Abelian case is
\begin{eqnarray}
\lefteqn{
{\cal E}_{\rm mf} = 
- \sum_\alpha \left\{ 4 [u(\Phi_\alpha) - \Phi_\alpha  u'(\Phi_\alpha)]
+ 6 \beta_\alpha [u'(\Phi_\alpha)]^4 \right\} } \nonumber \\
& & {} - \sum_\talpha \left\{ u(\Phi_\talpha) - \Phi_\talpha  u'(\Phi_\talpha) 
+ 4 \beta'_\talpha  \phi_\alpha \cdot \phi_{\alpha + 1}
u'(\Phi_\alpha) u'(\Phi_{\alpha+1}) [u'(\Phi_\talpha)]^2 \right\} .
\label{pot}
\end{eqnarray}
Here $\Phi_\alpha$ and $\Phi_\talpha$ are the mean fields
conjugate, respectively, to the link variables
$U_{\alpha\mu}$ and $U_\talpha$ and represented as two-dimensional
vectors, $\phi_\alpha$ is the unit vector in the direction of $\Phi_\alpha$,
a dot ($\cdot$) denotes the dot product of such vectors, and
$u(\Phi) = \ln I_0(\Phi)$, where $I_0$ is the modified Bessel function. The numerical
coefficients in \eq{pot} reflect the number of links (four) and plaquettes (six) per site
of a 4-dimensional hyperplane.

Suppose that in the region of interest $\beta_\alpha$ is large, in comparison with
both $\beta'_\talpha$ and unity. Then, we can first minimize the in-plane 
potential---the
first term in \eq{pot}---and use the obtained value of $\Phi_\alpha$ in the second,
interplane, term. We find that the in-plane field is large, 
$\Phi_\alpha \gg 1$, so that
\be
\langle U_{\alpha\mu} \rangle = u'(\Phi_\alpha) \approx 1 \; .
\label{uprime}
\ee
The interplane potential becomes
\be
{\cal E}'_{\rm mf} = 
- \sum_\talpha \left\{ u(\Phi_\talpha) - \Phi_\talpha  u'(\Phi_\talpha) 
+ 4 \beta'_\talpha  \phi_\alpha \cdot \phi_{\alpha + 1} [u'(\Phi_\talpha)]^2 \right\} .
\label{inter}
\ee
This is essentially the mean-field potential of the 4-dimensional ${\rm O}(2)$ 
nonlinear sigma model. We expect this theory to have a second-order phase transition
at some critical value $\beta'_\talpha = \beta'_{\rm cr}$. Indeed, 
at small $\Phi_\talpha$,
\be
{\cal E}'_{\rm mf} \approx \sum_\talpha  \left\{ \frac{1}{4} \Phi_\talpha^2 
- \frac{3}{64} \Phi_\talpha^4 - \beta'_\talpha \phi_\alpha \cdot \phi_{\alpha + 1}
\left( \Phi_\talpha^2 - \frac{1}{4} \Phi_\talpha^4 \right) \right\} \; ,
\label{inter2}
\ee
which shows that, at the mean-field level, there is a second-order transition at 
$\beta'_\talpha = 1/4$.

In the case of a spatially-varying $\beta'_\talpha$, the critical value will be reached 
at some $\talpha = \talpha_+$. The corresponding value of $z$ is $z =z_+$.
Propagation of charges in the region $0\leq z \leq z_+$ is 5-dimensional, as 
shown by the nonzero expectation value
\be
\langle U_\talpha \rangle = u'(\Phi_\talpha) \neq 0 \; .
\label{exp1}
\ee
On the other hand, outside this region, 
\be
\langle U_\talpha \rangle = 0 \; .
\label{exp2}
\ee
This is the layer phase of the Abelian theory \cite{Fu:1983ei,Fu:1984gj}. 
We interpret \eq{exp2} as 
the absence of coherent propagation of charges across many layers. It is in this 
sense that individual layers ``decouple'': a short-distance correlation
(e.g., between nearest neighbors) remains.

Monte Carlo studies confirm the presence of a second-order phase transition for
the ${\rm U}(1)$ gauge group \cite{Hulsebos:1994pa,Dimopoulos:2006qz}. 
The mean-field theory predicts a second-order transition 
also for ${\rm SU}(2)$, but not for ${\rm SU}(3)$, the case of main interest to us. 
In the case of ${\rm SU}(3)$, the mean field $\Phi$ is a $3\times 3$ matrix, and this 
allows for a trilinear term, proportional to ${\rm det} \Phi$, in the potential. This 
term drives the sigma-model transition first-order already at the mean-field level
\cite{Kogut:1981ez}.

For non-Abelian fields, the would-be layer phase acquires a mass gap via 
confinement. On the other
hand, by choosing $\beta_\alpha$ to be large enough, we can make the confinement 
length $\Lambda$
of the layers as large as we please. This means that, while for non-Abelian
groups there is no 
genuine layer phase in five dimensions, there can be a layered state with a large but 
finite confinement length \cite{Fu:1983ei}. 

Outside the mean-field theory, the nonzero expectation values
(\ref{uprime}) and (\ref{exp1}) cannot
survive without gauge fixing \cite{Elitzur:1975im}. We therefore should be looking
at a gauge-invariant correlator of the sigma-model variables, namely,
\be
G_\talpha(x) = 
\langle {\rm Tr} [U_\talpha(x)\ldots U_\talpha^\dagger(0) \ldots ] \rangle \, ,
\label{C}
\ee
where the trace is in the fundamental representation of ${\rm SU}(3)$, and
the dots stand for link variables (in the $x^\mu$ directions) taken along 
the shortest path connecting $x$ to the origin. Thus, the correlator is
a Wilson loop stretched in the $x^\mu$ directions. The corresponding
correlation length $\lambda_\talpha$ can be defined by 
\be
\int d^4 x G_\talpha(x) = \lambda_\talpha^4 \, .
\label{corr_length}
\ee
In the ``normal'' region, $\lambda_\talpha$ is large---we expect it to go to
infinity in the limit when $\Lambda_{\rm QCD}$ goes to zero.
On the other hand, in the layers, $\lambda_\talpha$ is at most
a few times $\Delta x$ (unless the sigma-model phase transition is very weakly 
first-order). This disparity means that the boundary between the weakly-coupled 
phase and the layers remains sharp even for a compact fifth dimension, despite
the absence of a genuine order parameter.

Now let us add quarks to the theory. \eq{exp2} shows that, in the layered state, 
these will preferentially propagate along the layers and therefore behave as ordinary
4-dimensional quarks. As a result, each layer has a copy of chiral symmetry that 
acts on quarks and is spontaneously broken by the quark condensates. 
The pattern of chiral symmetry breaking can then be described in the same way as
in QCD, namely, with the help of $N_f \times N_f$ unitary matrices ${\cal U}_\alpha$
(one per layer), governed by a chiral effective Lagrangian,\footnote{
In general, we denote by $\Lfour$ a Lagrangian density referring to the 4-dimensional
coordinate volume (so that the corresponding action is $\int \Lfour d^4x$).}
\be
\Lfour_{\rm chiral} =  - f^2 \sum_\alpha  \eta^{\mu\nu} 
{\rm Tr}  ( D_\mu {\cal U}_\alpha^\dagger D_\nu {\cal U}_\alpha ) \, .
\label{chiral}
\ee
Here $f \sim \Lambda$, and the covariant derivative includes the electroweak
gauge fields.
The sum extends over all hyperplanes that are in the layered state.
In our scenario, this chiral Lagrangian
is responsible for the masses of the $W$ and $Z$ bosons.

\section{Masses of $W$ and $Z$} \label{sec:masses}
The covariant derivative in \eq{chiral} is
\be
D_\mu {\cal U}_\alpha = 
\left( \partial_\mu +i A_\mu^a \frac{\tau^a}{2} \right) {\cal U}_\alpha -
i {\cal U}_\alpha B_\mu \frac{\tau^3}{2} \; ,
\label{covar}
\ee
where $A_\mu^a$ and $B_\mu$ are respectively the ${\rm SU}(2)_W$ and ${\rm U}(1)_Y$ 
gauge fields, and $\tau^a$ are the Pauli matrices.

In the absence of quark condensates, the lowest-energy mode of a gauge field 
is a constant and has zero mass.
This follows directly from the mode equation at zero 3-momentum. In a suitable gauge,
it reads
\be
- \partial_z (a^2 \partial_z A)  = E^2 A \; ,
\label{mode_eq}
\ee
where $A$ is any of $A_\mu^a$ and $B_\mu$, and $E$ is the energy (mass) of the mode.

If the mass term produced by the condensates is sufficiently small (the condition for
this will be discussed below), the leading corrections to the eigenvalues $E^2$ can be
found by first-order perturbation theory. For the lowest mode, this
means computing the
Lagrangian (\ref{chiral}) on constant fields. Setting ${\cal U} = 1$ and the gauge 
fields to constants turns \eq{chiral} into
\be
\Lfour_{\rm chiral} = 
- \half f^2 n_g N_{\rm layer} \left[ (A_\mu^1)^2 + (A_\mu^2)^2 
+ (A_\mu^3 - B_\mu)^2 \right] 
\; ,
\label{mass}
\ee
where $n_g$ is the number of quark doublets, $N_{\rm layer}$ is the number
of hyperplanes in the layered region, and $(A_\mu)^2 \equiv \eta^{\mu\nu} A_\mu A_\nu$. 
This is the same as the mass Lagrangian in the minimal
standard model, provided we identify the symmetry-breaking scale $v \sim 250$ GeV as
\be
v^2 = 4 f^2 n_g N_{\rm layer} \; .
\label{v2}
\ee
We see that, for a large $N_{\rm layer}$, the parameter $f$ and, hence,
the confinement scale $\Lambda$ are much smaller than $v$.

To obtain the condition of applicability of \eq{mass}, consider how much the
lowest mode is deformed by the condensates. The mode equation now reads
\be
- \partial_z (a^2 \partial_z A)  + M^2(z) A = E^2 A \; ,
\label{with_mass}
\ee
where $M(z)$ is a constant, $M_0$,
at $z_+ < z \leq \pi R$ and zero otherwise. Details of the solution depend
on the form of the warp factor. 
We will describe results for
\be
a(z) = e^{-\kappa |z|} \; ,
\label{ads}
\ee
which represents a region of the anti-de Sitter (AdS) space bounded by the branes
\cite{Randall:1999vf}. In this case, expanding \eq{with_mass} 
in small $M_0$ and $E$, we obtain, to the leading nontrivial order,
\ba
A(y) & = & 1 - \half E^2 y^2 \left (\ln\frac{y}{y_0} - \half \right) ,
\hspace{7em} y_0 < y < y_+ \, , \label{sol1} \\
A(y) & = & C \left\{
1 + \half (M_0^2 - E^2) y^2 \left (\ln\frac{y}{y_R} - \half \right) \right\} ,
\hspace{1em} y_+ < y < y_R \, , \label{sol2}
\ea
where $y = \kappa^{-1} e^{\kappa z}$ (so that $z=0$ maps to 
$y_0 = \kappa^{-1}$, $z=z_+$ to $y = y_+$, and $z=\pi R$ to $y= y_R$) and
$C$ is a constant. Matching \eqs{sol1}{sol2} at $y = y_+$, we find
\be
E^2 = M_0^2 \left( 1 - \frac{z_+}{\pi R} \right) \; ,
\label{E2}
\ee
which is the same as $E^2$ obtained by using \eq{mass}. 
This value of $E$ is identified with the mass of $W$ or $Z$, depending on which
value of $M_0$ is used in \eq{with_mass}.
The requisite condition of applicability
is that the variation of $A(y)$, caused by the 
corrections in \eqs{sol1}{sol2}, is small compared to unity. For example, if
$z_+ \sim \pi R$, the condition becomes $m_W \ll \kappa e^{-\pi \kappa R}$.

\section{Fermions} \label{sec:ferm}
Propagation of quarks is 5-dimensional in the ``normal''  region $0 \leq z \leq z_+$ 
but, as we have seen in Sec.~\ref{sec:struc}, becomes 4-dimensional in the layered 
state. Because of the twisted boundary conditions (discussed below), it is
convenient to consider first propagation of quarks with the ``normal'' region
unfolded into $z_- \leq z \leq z_+$ (see Fig.~\ref{fig:hyper})
and impose the $Z_2$ symmetry later.
Thus, we begin with the 5-dimensional Dirac equation supplemented by
boundary conditions at $z=z_\pm$. These boundary conditions encode the way in
which the quarks are reflected from the layered region.

For computation of the quark masses,
it is sufficient to consider solutions that are independent 
of the spatial coordinates $x^i$. Then, the fermions can be assumed two-component,
and the $\gamma$-matrices two-dimensional.
We use the representation in which 
$\gamma^0 = \sigma^1$ and $\gamma^z = -i \sigma^2$, where $\sigma^a$ are the Pauli
matrices. 

Let us first neglect the penetration of ``normal'' quarks into the layered region.
Then, the space for them ends at $\alpha = \alpha_\pm$, and the lattice 
Lagrangian of a single free quark is
\be
\Lfour_F = 
\sum_{\alpha = \alpha_-}^{\alpha_+} i w_\alpha 
\psi_\alpha^\dagger \partial_t \psi_\alpha 
+ \beta_F  \sum_{\alpha = \alpha_-}^{\alpha_+ - 1}
(i \psi_\alpha^\dagger \sigma^3 \psi_{\alpha+1} + \mbox{H.c.} ) \, ,
\label{Lf_lat}
\ee
with some weights $w_\alpha > 0$ and $\beta_F \equiv 1/2 \Delta z$. (We will 
discuss shortly how this is related to the continuum Lagrangian.)
For $\psi_\alpha$ depending 
on time as $e^{-iEt}$, the discrete Dirac equation following from \eq{Lf_lat} reads
\be
- i \beta_F \sigma^3 (\psi_{\alpha + 1} - \psi_{\alpha -1}) = E w_\alpha \psi_\alpha
\label{dirac_lat}
\ee
at the interior points, and
\ba
- i \beta_F \sigma^3 \psi_{\alpha_- + 1} & = & E w_{\alpha_-} \psi_{\alpha_-} \; , 
\label{dirac-} \\
i \beta_F \sigma^3 \psi_{\alpha_+ - 1} & = & E w_{\alpha_+} \psi_{\alpha_+} 
\label{dirac+}
\ea
at the ends. 

Five-dimensional fermions hop between the 4-dimensional hyperplanes 
in such a way that, if a fermion is left-handed (in the 4-dimensional sense) 
at site $\alpha$, it will be right-handed at $\alpha +1$, and vice versa.
As a consequence, existence of purely chiral (left- or right-handed) modes depends
on whether the total number of sites in the ``normal'' region is even or odd. 
Here we consider the case when it is odd, and
such modes exist (the logic being that we are looking for a discretization
that produces phenomenologically viable spectra).

In this case, both $\alpha_+$ and $\alpha_-$
can be made even (by relabeling the sites if necessary), and
the system (\ref{dirac_lat})--(\ref{dirac+}) has a $E=0$ solution:
$\psi_\alpha=\psi_0$ 
(a constant nonzero spinor) at even $\alpha$, and $\psi_\alpha=0$ at odd.
Decomposing $\psi_0$ into left- and right-handed components, we obtain two purely
chiral solutions---one left- and one right-handed.

Now, let us take into account the mixing of the left- and right-handed species
due to quark condensation in the layered region.\footnote{
The way it was interpreted
in Sec.\ref{sec:struc}, the condition (\ref{exp2}) means that there is no coherent
propagation of quarks across many layers. The interlayer correlation length, 
however, while short, is finite, so 
the ``normal'' quarks extend somewhat into the layered region.}
The effect is most easily computed directly in the continuum limit.
Consider two continuum
5-dimensional fermions $\Psi_1$ and $\Psi_2$, Combining them into a single
column, called $\Psi$, we can write the kinetic term in the action as
\be
S_F = i \int d^4 x dz \sqrt{-g}
\bar{\Psi} \gamma^A e_{A}^{~N} (\partial_N + \Gamma_N) \Psi \; ,
\label{Lf}
\ee
where $e_{A}^{~N}$ are basis vectors, and $\Gamma_N$ is the spin connection
($A,N=0,\ldots, 4$). For the metric (\ref{met}), the 
nontrivial entries are
\ba
e_\alpha^{~\nu} & = & \frac{1}{a} \delta^\nu_\alpha \; , \\
\Gamma_\nu & = & - \half a' \eta_{\nu\alpha} \gamma^\alpha \gamma^z 
\ea
($\alpha,\nu = 0,\ldots,3$). The Dirac equation then reads
\be
- i \gamma^0 \left[
\gamma^i \partial_i + a \gamma^z \left(
\partial_z +  \frac{2 a'}{a} \right) \right] \Psi = E \Psi \; .
\label{Dirac}
\ee
Thus, the effect of the spin connection can be absorbed by redefinition of 
the field into $\psi = \Psi a^2$. In the continuum problem, it is convenient 
to make a further change of variables---from
$z$ to
\be
\eta = \int_0^z \frac{dz'}{a(z')} \, .
\label{eta}
\ee
These redefinitions bring \eq{Dirac} to the usual flat-space form.

As in the lattice version, we concentrate on solutions that are independent 
of $x^i$. Then, $\psi_1$ and $\psi_2$ can each be assumed two-component, and
\eq{Dirac} becomes
\be
- i \sigma^3 \partial_\eta \psi = E \psi \; .
\label{Dirac2}
\ee
Here $\psi=(\psi_1, \psi_2)^T$ has the total of four components, and
$\sigma^3$ acts individually on $\psi_1$ and $\psi_2$.
For each of those, \eq{Dirac2} is a continuum version of 
\eq{dirac_lat} with $w_\alpha = 1/ a(z_\alpha)$.

To reproduce the chiral spectrum obtained on the lattice, we choose the
boundary conditions at $\eta = \eta_\pm$
in such a way that, in the absence of any mixing between the species,
one of $\psi_{1,2}$ would have a purely left-handed and the other a purely
right-handed zero-energy mode. This is conveniently achieved by extending the range 
of $\eta$ to the entire infinite line  
and adding a mass term that produces chiral modes in the region 
$\eta_- \leq \eta \leq \eta_+$:\footnote{Of course, only discrete eigenstates
of this extended problem are relevant.}
\be
{\cal L}_M = - M(\eta) (\bar{\psi}_1 \psi_1 - \bar{\psi}_2 \psi_2 ) \; ,
\label{LM}
\ee
where 
\be
M(\eta) = \left\{ \begin{array}{cc} M_-, & \eta < \eta_-, \\
0, & \eta_- < \eta < \eta_+, \\
M_+, & \eta < \eta_+, \end{array} \right.
\label{Meta}
\ee
and $M_-$ and $M_+$ are of opposite signs. This mechanism of producing chiral
fermions is similar to the domain-wall mechanism of Ref.~\cite{Rubakov:1983bb}, 
but instead of localizing
fermions on a domain wall \cite{Jackiw:1975fn} it confines them to the entire
``normal'' region.
Condensation of quarks in the layered region is represented by the off-diagonal term
\be
{\cal L}_m = - \mu(\eta) \bar{\psi}_1 \psi_2 - \mu^*(\eta) \bar{\psi}_2 \psi_1 \; ,
\label{Lm}
\ee
where $\mu(\eta)$ has the same form as \eq{Meta} but with asymptotic values 
$\mu_-$ and $\mu_+$. The total mass Lagrangian density is ${\cal L}_M + {\cal L}_m$.

We wish to stress 
that adding these mass terms is merely a convenient way to describe scattering of 
quarks
off the layered region. Indeed, all the mass parameters appearing in \eqs{LM}{Lm} 
can in the end be taken to infinity; all that matters is their ratios.

The mass matrix of $\psi_1$ and $\psi_2$ (we will refer to these components as
``doublers'') can be diagonalized at $\eta > \eta_+$ by a unitary transformation 
involving only the generators $\tau^1$ and $\tau^2$ of the doubler 
${\rm SU}(2)$:\footnote{We use a different notation to distinguish these from the 
$\sigma^a$ matrices that act on the spinor index.}
\be
\left( \begin{array}{cc} M_+ & \mu_+ \\
\mu_+^* & - M_+ 
\end{array} \right) = e^{i\theta^j_+ \tau^j} 
\left( \begin{array}{cc} \tM_+ & 0 \\
0 & - \tM_+ 
\end{array} \right) e^{-i\theta^j_+ \tau^j} \, ,
\label{diag}
\ee
where $j=1,2$ and $\tM_+^2 = M_+^2 + |\mu_+|^2$. A similar transformation 
with parameters $\theta^j_-$ diagonalizes the mass matrix at $\eta < \eta_-$.
In general, there is no reason why the parameters of these two transformations
should be the same: 
the mass matrices in the two regions can be misaligned. Such a misalignment 
is analogous to a difference in the phase of the superconducting order parameter 
on two sides of a Josephson junction. By this analogy, we expect it to lead
to persistent vacuum currents in the $z$ direction, $\langle J^z \rangle \neq 0$,
where
\be
J^z = {\bar \psi} \gamma^z T \psi \, ,
\label{Jz}
\ee
and $T$ is $\tau^1$ or $\tau^2$.

It is possible to define the orbifold projection of fermions
so as to preserve nonzero expectation values
of these currents. Indeed, define the projection as 
invariance under $\psi(\eta) \to Z \psi(-\eta)$, where
\be
Z = i \tau^3 \gamma^z 
\label{Z}
\ee
(with $\tau^3$ acting in the doubler space).
The current (\ref{Jz}) is even under this transformation provided\footnote{
This is apparently the same as the consistency condition for a twist in the 
Scherk-Schwarz mechanism \cite{Scherk:1978ta,Scherk:1979zr} of symmetry breaking 
(see, for example, Ref.~\cite{Bagger:2001qi}).}
\be
ZTZ = - T \, ,
\label{ZTZ}
\ee
which holds for both of the $\tau$ matrices in question.

The mass term $L_M$ is even under the orbifold transformation when
$M_+ = - M_-$, while $L_m$ is even when $\mu_+ = \mu_-$ (no minus sign here!).
Under these conditions, any nonzero $\mu_+$ leads to a misalignment of the mass
matrices. Let us show that such a misalignment gives rise to a nonzero quark mass. 
When more quark flavors are added, misalignments in the flavor space
will similarly break flavor symmetries.

In the layered region, the relevant wavefunctions are purely evanescent:
\ba
\psi(\eta < \eta_-) & = & \left\{ c_1 
\left( \begin{array}{c} E-i\tE_- \\ M_- \\ 0 \\ \mu_-^* \end{array} \right)
+ c_2 
\left( \begin{array}{c} 0  \\ \mu_- \\ E-i\tE_- \\ -M_- \end{array} \right) 
\right\} e^{\tE_- (\eta -\eta_-)} ,
\label{evan-} \\
\psi(\eta > \eta_+) & = & \left\{ c_3
\left( \begin{array}{c} E+i\tE_+ \\ M_+ \\ 0 \\ \mu_+^* \end{array} \right)
+ c_4
\left( \begin{array}{c} 0  \\ \mu_+ \\ E+i\tE_+ \\ -M_+ \end{array} \right) 
\right\} e^{-\tE_+(\eta - \eta_+)} ,
\label{evan+}
\ea
where $\tE_\pm = [\tM_\pm^2  - E^2]^{1/2}$,
and $c_i$ are constants. Matching these evanescent solutions 
to the solution of \eq{Dirac2} for the interior region, we obtain the ratios of 
the constants and an equation for the eigenvalues:
\be
2 \tM_- \tM_+ \cos[2EL + \delta(E)] = 2 M_+ M_- + \mu_-^* \mu_+ + \mu_+^* \mu_- \; ,
\label{eigen}
\ee
where $\delta(E)$ is a phase, defined as follows:
\be
\frac{E-i\tE_-}{E+i\tE_+} = \frac{\tM_-}{\tM_+} e^{i\delta(E)} \; ,
\label{delta}
\ee
and
\be
L = \eta_+ - \eta_- = \int_{z_-}^{z_+} \frac{dz}{a(z)} \; .
\label{length}
\ee
(Recall that $M_+$ and $M_-$ are of opposite signs.)

In the orbifolded case, $M_+ = -M_- \equiv M$ and $\mu_+ = \mu_- \equiv \mu$.
When $|\mu|, 1/L \ll |M|$, there are light states, for which $|E| \ll \tM_\pm$, 
i.e., $\delta \approx -\pi$. Setting
$\delta = -\pi$ leads to the following spectrum:
\be
E_n = E_0 + \frac{\pi}{L} n \; ,
\label{En}
\ee
where $E_0 \ll 1/L$, and $n\geq 0$ is an integer. We identify $E_0$ with the mass
of a standard-model quark. 

A sufficiently small $L$ will render the modes
with $n\geq 1$ unobservable. For example, for the AdS case (\ref{ads}),
\be
L = \frac{2}{\kappa}  \left( e^{\kappa z_+}  - 1 \right) \; .
\label{Lads}
\ee 
The condition that $E_1$ is much larger than the weak scale $m_W$ becomes
$\kappa e^{-\kappa z_+} \gg m_W$.

\section{Conclusion} \label{sec:concl}
In this paper, we explored the possibility that, by suitably discretizing a warped
fifth dimension, one obtains a bulk region in which quantum chromodynamics is
in a layered state, and the proximity effect due to the presence
of such a region reproduces the physics of the standard
model. In our scenario, the layered state is characterized by technicolor-like
confining dynamics, but instead of technicolor we have the usual color,
which is stronger in the layers than in the ``normal'' world.

Both the vector bosons and the quarks acquire masses by scattering off the layered
region. The ``normal'' quarks couple appreciably
only to the outermost of the layers, but
the $W$ and $Z$ to all of them. As a result, the
confinement scale of the layer theories can be much lower than the conventional
$v\sim 250$ GeV, with potentially interesting phenomenological consequences.

The mechanism described here produces masses for the quarks but not for the 
leptons. Can
masses of the leptons be produced in a similar way? One possibility is that 
electromagnetism becomes strongly-coupled in the bulk and forms a layer phase
of its own (a genuine one, as the theory is Abelian). In the layer phase,
hopping of individual electrons between the layers is inhibited, but 
an electron-positron pair, being neutral, can hop freely. As a result, there
is an effective short-distance attraction between opposite charges, which  
may be strong enough to cause condensation in the $\bar{e}_L e_R$ 
channel. Such condensates would give masses to all the charged leptons while 
leaving the neutrinos massless.

\acknowledgments
The author thanks T. Clark, S. Love, and M. Shaposhnikov for discussions of the 
results.
This work was supported in part by the U.S. Department of Energy through grant 
number DE-FG02-91ER40681.

\end{document}